\documentclass[twocolumn,showpacs,preprintnumbers,amsmath,amssymb,pra]{revtex4}

\usepackage{graphicx}
\usepackage{dcolumn}
\usepackage{bm}

\begin{document}

\preprint{APS/123-QED}

\title{Spontaneous Radiation and Amplification of Kelvin Waves on Quantized Vortices in Bose-Einstein Condensates}

\author{Hiromitsu Takeuchi,${}^1$ Kenichi Kasamatsu,${}^2$ and Makoto Tsubota${}^1$}
\affiliation{%
${}^1$Department of Physics, Osaka City University, Sumiyoshi-ku, Osaka 558-8585, Japan \\
${}^2$Department of Physics, Kinki University, Higashi-Osaka, Osaka 577-8502, Japan
}%
\date{\today}

\begin{abstract}
We propose a different type of Landau instability in trapped Bose-Einstein condensates by a helically moving environment.
In the presence of quantized vortices, the instability can cause spontaneous radiation and amplification of Kelvin waves.
This study gives a microscopic understanding of the Donnelly-Glaberson instability which was known as a hydrodynamic instability in superfluid helium.
 The Donnelly-Glaberson instability can be a powerful tool for observing the dispersion relation of Kelvin waves, vortex reconnections, and quantum turbulence in atomic Bose-Einstein condensates.
\end{abstract}

\maketitle

\section{INTRODUCTION}
 Even if an external environment, such as a container wall, is moving in the laboratory frame, superfluid component can remain at rest in thermal equilibrium, which is frictionless flow.
 The thermodynamic stability of the frictionless flow is discussed with the thermodynamic energy,
\begin{equation}
E({\bm V},{\bm \Omega})=E_0-{\bm V}\cdot {\bm P}-{\bm \Omega}\cdot{\bm L},
\label{eq:thermodynamic}
\end{equation}
 where $E_0$, ${\bm P}$ and ${\bm L}$ are the energy, the momentum, and the angular momentum of the superfluid in the laboratory frame, respectively.
 Here, ${\bm V}$ and ${\bm \Omega}$ refer to the linear velocity and the rotational frequency of the motion of the environment, respectively.
 The second term on the right-hand side of Eq. (\ref{eq:thermodynamic}) is essential to the discussion for superfluid although it is usually neglected for ordinary liquid \cite{Landau}.
 The frictionless flows are realized at local minima of the thermodynamic energy of Eq. (\ref{eq:thermodynamic}).
 This kind of thermodynamic consideration has been applied only to the translational (${\bm V}\neq 0$ and ${\bm \Omega}=0$) and the rotational (${\bm V}=0$ and ${\bm \Omega}\neq 0$) cases, which has been thoroughly studied in superfluid helium and atomic Bose-Einstein condensates (BECs) \cite{PitaevskiiStringari}.
 The former was applied to the famous discussion of the Landau critical velocity for bulk superfluid.
 The latter was discussed for understanding the anomalous moment of inertia and the vortex lattice formation.
 The coupled situation (${\bm V}\neq 0$ and ${\bm \Omega}\neq 0$), which has not been known well to date, reveals a new aspect of superfluidity.

 In this work, we theoretically study the superfluidity under a helically moving environment with ${\bm V} \parallel {\bm \Omega}$ in the presence of quantized vortices in atomic BECs.
 It is revealed that helical vortex modes (Kelvin waves) \cite{Kelvin} can be spontaneously radiated and amplified due to the Landau instability.
 From the viewpoint of thermodynamics,
 this phenomena could be called the generalized Donnelly-Glaberson (DG) instability,
 which was phenomenologically understood as hydrodynamic instability causing the amplification of Kelvin waves on vortex lines in superfluid helium \cite{Glaberson,DonnellyBook}.

 This paper is organized as follows.
 First, we introduce the DG instability in superfluid helium with a phenomenological model to compare with the instability in atomic BECs.
 Next, the DG instability is microscopically discussed in atomic BECs with the Gross-Pitaevskii (GP) and the Bogoliubov-de Gennes (BdG) models.
 Finally, it is shown that the DG instability can be applied to the observation of the dispersion relation of Kelvin waves, vortex reconnections \cite{Ogawa}, and quantum turbulence \cite{QT, TsubotaJPSJ} in atomic BECs.

\section{DONNELLY-GLABERSON INSTABILITY IN SUPERFLUID HELIUM}
 We now introduce an intuitive description of the DG instability in the simplest case of axial normal flow along an isolated vortex line under the vortex filament model in superfluid helium \cite{DonnellyBook}.
 When the line is deformed into a helix with the radius $\epsilon$, the wave vector $k$, and the angular velocity $\omega$, the position ${\bm s}$ of the line may be parametrized with $\xi$ as ${\bm s}(\xi,t)=\epsilon \cos(kz(\xi)-\omega t)\hat{\bm x}+\epsilon \sin(kz(\xi)-\omega t)\hat{\bm y}+z(\xi)\hat{\bm z}$.
 In the localized induction approximation which neglects interactions between vortices \cite{DonnellyBook}, the equation of motion of vortex lines is written as
 $d{\bm s}/dt={\bm v}_{i}+\alpha {\bm s}^1 \times ({\bm v}_{N}-{\bm v}_{i})$,
 where we use ${\bm s}^n=d^n{\bm s}/d{\xi}^n$, the mutual friction coefficient $\alpha(T) \geq 0$, the normal fluid velocity ${\bm v}_{N}$, and the local self-induced velocity ${\bm v}_{i}$.
 When $\epsilon k \ll 1$,  the self-induced velocity is linearized to ${\bm v}_{i} = \beta {\bm s}^1 \times {\bm s}^2 \approx \beta k^2\epsilon \sin(kz(\xi)-\omega t)\hat{\bm x}-\beta k^2 \epsilon \cos(kz(\xi)-\omega t)\hat{\bm y}$ with $\beta=\frac{\kappa}{4\pi} \ln(\frac{1}{ka})$, the circulation quantum $\kappa$, and vortex core radius $a$.
 If the normal component is negligible at $T=0$, $\alpha(T=0)=0$, the Kelvin wave propagates keeping its initial configuration and rotating with frequency $\omega_0 = \beta k^2$.
  At finite temperatures under the helical normal-fluid flow with ${\bm v}_{N}=\Omega \hat{\bm z} \times {\bm r}+V\hat{\bm z}$, where $\Omega$ and $V$ are positive constants, the initial stage of the dynamics is governed by
\begin{equation}
\frac{d \epsilon}{dt} =-\alpha(\omega_0+\Omega-kV)\epsilon.
\label{eq:DG_He}
\end{equation}
 The radius of the helix with $k$ is increased or decreased with time for $\omega_0+\Omega-kV<0$ or $>0$, respectively.
 Therefore, the straight vortex line becomes unstable when the velocity $V$ exceeds the DG critical velocity
\begin{equation}
 V_{\rm DG}=\min_k\Bigl( \frac{\omega_0+\Omega}{k}\Bigr).
\label{eq:V_DG}
\end{equation} 
 Since the instability is determined by the local configuration of the vortex lines, a similar mechanism can be applied to each vortex line in more complicated cases such as vortex lattices and vortex tangles by considering the total velocity field originating from all vortices beyond the localized induction approximation \cite{Tsubota_Araki,Finne}.
 Thus, the DG instability plays an important role in the vortex dynamics in superfluid ${}^4$He and ${}^3$He-B at finite temperatures.

\section{THERMODYNAMIC APPROACH}
Let us generalize the DG instability from a thermodynamic point of view, which is not clear in the vortex-filament approach.
 If we consider the normal-fluid component with ${\bm v}_{N}=\Omega \hat{\bm z} \times {\bm r}+V\hat{\bm z}$ as an environment for the system,
 we have the thermodynamic energy $E(V,\Omega)=E_0-VP_z-\Omega L_z$ for the superfluid component from Eq. (\ref{eq:thermodynamic}).
Here, $P_z$ and $L_z$ are the momentum and the angular momentum of the superfluid component along the $z$-axis, respectively.
 Note that, in general, the role of the environment can be played by some other objects such as a container wall or something, instead of the normal component.
 Thus the discussion is also applicable to the case of zero temperature through this generalization.
 For a given $\Omega$, we expect that the stable states with straight vortex lines are realized at the local minima in $E(V,\Omega)$ below the DG critical velocity $V_{\rm DG}$.
 Then, we suppose that the thermodynamic instability associated with disappearance of the local minimum leads to the onset of the amplification of Kelvin waves in the DG instability above $V_{\rm DG}$.
This kind of thermodynamic instability can be explained microscopically by the Landau instability,
 where the energies of elementary excitations become negative in a co-moving frame of the environment.
Then, spontaneous radiation and amplification of these elementary excitations occur to decrease the thermodynamic energy.

However, it is difficult to analyze the DG instability from the above view point in superfluid helium systems,
 where the vortex dynamics was analyzed with the phenomenological vortex filament model.
This presents a contrast with atomic BEC systems,
where calculations from first principles give exact solutions of vortex states and make it possible to discuss Kelvin waves microscopically as elementary excitations of the system \cite{Pitaevskii, Mizushima, Fetter, Simula}.
From the discussion in atomic BECs, we can obtain the microscopic origin of the DG instability as {\it spontaneous radiation and amplification of Kelvin waves due to the Landau instability}.

\section{DONNELLY-GLABERSON INSTABILITY IN ATOMIC BOSE-EINSTEIN CONDENSATES}
 Let us consider quantized vortices in trapped BECs.
 For simplicity we assume a periodic system along the rotation axis and use an axisymmetric harmonic potential $V_{\rm t}({\bm r})=M\omega_{t}^2\rho^2/2$ with cylindrical coordinates $(\rho,\theta,z)$, where $M$ and $\omega_{t}$ are the atomic mass and trapping frequency, respectively.
 The thermodynamic energy for the superfluid component is described by the macroscopic wave function $\Psi$.
 Under the constraint that the total particle number $N$ in the system is constant, in the helically moving frame we have the thermodynamic energy
\begin{eqnarray}
K(\Psi,V,\Omega)=K_0(\Psi)-VP_z(\Psi)-\Omega L_z(\Psi),
\label{eq:Kpotential}
\end{eqnarray}
 with $K_0(\Psi)=\int d{\bm r} [\Psi^*( -\frac{1}{2}\nabla^2+\frac{1}{2}\rho^2+g|\Psi|^2 -\mu)\Psi]$, $P_z(\Psi)=\int d{\bm r} \Psi^*\hat{p}_z\Psi$, and $L_z(\Psi)=\int d{\bm r} \Psi^*\hat{l}_z\Psi$, where $\mu$ is the chemical potential and we used $\hat{p}_z=-i{\nabla}_z$ and $\hat{l}_z=-i\left({\bm r} \times {\bm \nabla}\right)_z$. Here, the units of energy, length, and time are given by the corresponding scales of the harmonic potential as $\hbar\omega_{t}$, $b_{\bot}=\sqrt{\hbar/M\omega_{t}}$, and $\omega_{t}^{-1}$, respectively.
 The wave function is normalized as $\int d\rho\rho\int d\theta\int^{L/2}_{-L/2} dz|\Psi({\bm r})|^2=1$, where $L$ is the periodicity along the $z$-axis.
 The atomic interaction is characterized by $g$, which is proportional to the $s$-wave scattering length $a$ as $g=4\pi aN/b_{\bot}>0$.
 From the potential of Eq. (\ref{eq:Kpotential}), we obtain the time-dependent GP equation
\begin{eqnarray}
{\textstyle
i\frac{\partial}{\partial t}\Psi= \bigl( -\frac{1}{2}\nabla^2+\frac{1}{2}\rho^2+g|\Psi|^2 -\mu-V\hat{p}_z -\Omega\hat{l}_z  \bigr) \Psi.
}
\label{eq:GP}
\end{eqnarray}

\subsection{Excitations in the single-vortex state}
 First, we discuss the simplest case of a straight vortex line located along the $z$-axis.
 The wave function $\Psi_0$ in the stationary state can be written in an axisymmetric form as $\Psi_0=\psi_0(\rho)e^{i\theta}$.
 Due to the symmetry of the wave function, the chemical potential $\mu$ in the co-moving frame may be defined as
\begin{eqnarray}
\mu \equiv \mu(V,\Omega)=\mu(0,0)-\Omega;
\label{eq:mu}
\end{eqnarray}
 $\psi_0$ is then independent of $V$ and $\Omega$.

 We now represent the collective modes with the perturbed wave function $\Psi =\Psi_0+\delta\Psi$.
 Then, a collective excitation with frequency $\omega$ is written as
\begin{eqnarray}
\delta \Psi=e^{i\theta}[u_{k,l}(\rho)e^{i(kz+l\theta-\omega t)}-v^*_{k,l}e^{-i(kz+l\theta-\omega t)}],
\label{eq:perturb}
\end{eqnarray}
 where $k$ and $l$ refer to the wave number and the angular quantum number of the excitation along the $z$-axis, respectively.
 Here, we consider only the lowest modes along the radial direction. The normalization is $\int dV(u_{k,l}^*u_{k',l'}-v_{k',l'}^*v_{k,l})=\eta \delta_{k,k'}\delta_{l,l'}$, where $\eta>0$ and $\delta_{i,i'}$ is the Kronecker delta.

\begin{figure}[htbp]
  \includegraphics[width=1. \linewidth]{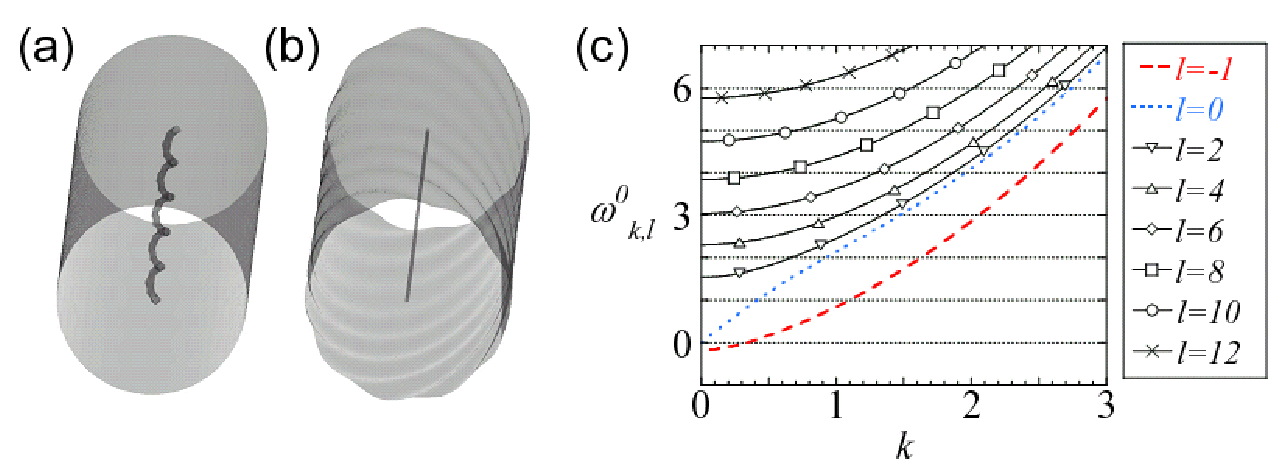}
\caption{
 Constant surface densities $|\Psi|^2$ of condensates with (a) a Kelvin wave and (b) a surface wave, which are obtained by numerically solving Eq. (\ref{eq:GP_gamma}) for $(V,\Omega)=(1.071,0.35)$ and $(V,\Omega)=(0.8,0.45)$, respectively.
 (c) Dispersion relation $\omega^0_{k,l}=\omega_{k,l}(V=0,\Omega=0)$ of Kelvin waves with $l=-1$ (dashed line), varicose waves with $l=0$ (dotted line), and several surface waves with $l>0$ (solid lines) for the single-vortex state with $g_{\rm 2D} = 500$.
}
\label{fig:spect}
\end{figure}

 The lowest modes can be classified into three groups by the angular quantum number $l$.
 One group is Kelvin waves with $l=-1$, which deforms the vortex line into a helix \cite{Pitaevskii}.
The second is density waves with $l=0$, which propagate keeping the condensate density axisymmetric.
 We call this mode the ``varicose wave'' analog to classical fluid, for which the core diameter of a vortex varies as the wave propagates \cite{DonnellyBook}.
 The last group consists of surface waves with $l \neq -1$ and $0$, which disturb the condensate density only near the surface.
 Typical Kelvin and surface waves are shown in Figs. \ref{fig:spect}(a) and \ref{fig:spect}(b).

 Linearizing the GP Eq. (\ref{eq:GP}) with Eq. (\ref{eq:perturb}), we obtain the BdG equations
\begin{eqnarray}
(\omega+kV+l\Omega) \left( 
\begin{array}{cc}
u_{k,l} \\
v_{k,l} \\
\end{array} 
\right)=\left( 
\begin{array}{cc}
\hat{h}_+ & -g\psi_0^2 \\
g{\psi_0^*}^2 & -\hat{h}_- \\
\end{array} 
\right) \left( 
\begin{array}{cc}
u_{k,l} \\
v_{k,l} \\
\end{array} 
\right),
\label{eq:BdG_ax}
\end{eqnarray}
where
$\hat{h}_{\pm}=-\frac{1}{2}\bigl[ (d^2/d\rho^2)+(d/\rho d\rho)-((l \pm 1)^2/\rho^2)-k^2 \bigr]+\rho^2/2+2g|\psi_0|^2-\mu(0,0)$.
 Analogous to Eq. (\ref{eq:mu}), we can define the frequency $\omega$ in the co-moving frame as
\begin{eqnarray}
\omega \equiv \omega_{k,l}(V,\Omega) =\omega^0_{k,l}-kV-l\Omega,
\label{eq:omega_kl}
\end{eqnarray}
 where $\omega^0_{k,l}\equiv \omega_{k,l}(0,0)$ is the dispersion relation in the laboratory frame.
 Then, $u_{k,l}$ and $v_{k,l}$ are independent of $V$ and $\Omega$.
 Figure \ref{fig:spect}(c) shows the dispersion relation $\omega^0_{k,l}$ for various $l$ by numerically solving Eqs. (\ref{eq:GP}) and (\ref{eq:BdG_ax}) for $g_{\rm 2D}\equiv g/L=500$.

\subsection{Landau critical velocity}
 If there is at least one mode with $\omega<0$, the stationary state $\Psi_0$ is a saddle point of the thermodynamic energy of Eq. (\ref{eq:Kpotential}).
 Then, the state becomes thermodynamically unstable due to the Landau instability and the mode should be spontaneously radiated and amplified to decrease the thermodynamic energy of Eq. (\ref{eq:Kpotential}).
 The stability of the single-vortex states for $V=0$ was investigated in Ref. \cite{IsoshimaCritical}.
 If the rotational frequency $\Omega$ is smaller than a critical value $\Omega_{L}$ for $V=0$,
 the single-vortex states are unstable and changed into the vortex-free state.
 Then, a Kelvin wave with $k=0$ is preferred to be excited,
 where the vortex core is transferred from the center $\rho=0$ so that the angular momentum of the condensate is decreased.
 Thus the lower critical frequency $\Omega_{L}=-\omega^0_{0,-1}$ is obtained from $\omega_{0,-1}(0,\Omega_{L})=0$ of Eq. (\ref{eq:omega_kl}).
 There is also the upper critical frequency $\Omega_{U}$.
 When $\Omega$ exceeds $\Omega_{U}$,
 some surface waves are excited in order to drag vortices into the condensate to increase its angular momentum.
 Then we have $\Omega_{U}\equiv \min_l\left(\omega^0_{0,l}/l\right)~(l >0)$ from Eq. (\ref{eq:omega_kl}).
 Thus the single-vortex states can be stabilized when $\Omega_{L} < \Omega <\Omega_{U}$.
 However, even if $\Omega_{L} < \Omega <\Omega_{U}$,
 the single-vortex states become unstable in the presence of $V$.
 The frequency $\omega_{k,l}(V,\Omega)$ of the mode with $l$ becomes negative when $V$ exceeds the Landau critical velocity
\begin{eqnarray}
V_l(\Omega)
=\min _k \biggl(\frac{\omega^0_{k,l}-l\Omega}{k}\biggr)
=\frac{\omega^0_{k_l,l}-l\Omega}{k_l},
\label{eq:V_l}
\end{eqnarray}
 where $k_l$ is the critical wave number.
 Then, the mode with the angular quantum number $l$ and the wave number $k_l$ can be spontaneously radiated due to the Landau instability.
 In particular, Kelvin waves with wave number $k_{-1}$ are spontaneously radiated when $V$ exceeds $V_{-1}$,
 which  represents the onset of the DG instability due to the Landau instability.
 In fact, we see that the critical velocity $V_{-1}$ for Kelvin waves has the same form as the DG criterion of Eq. (\ref{eq:V_DG}) in the vortex filament model.
 In this way, we microscopically derived the DG criterion from the thermodynamic consideration.

\subsection{Condition for the Donnelly-Glaberson instability}
\begin{figure} [htbp] \centering
  \includegraphics[width=1. \linewidth]{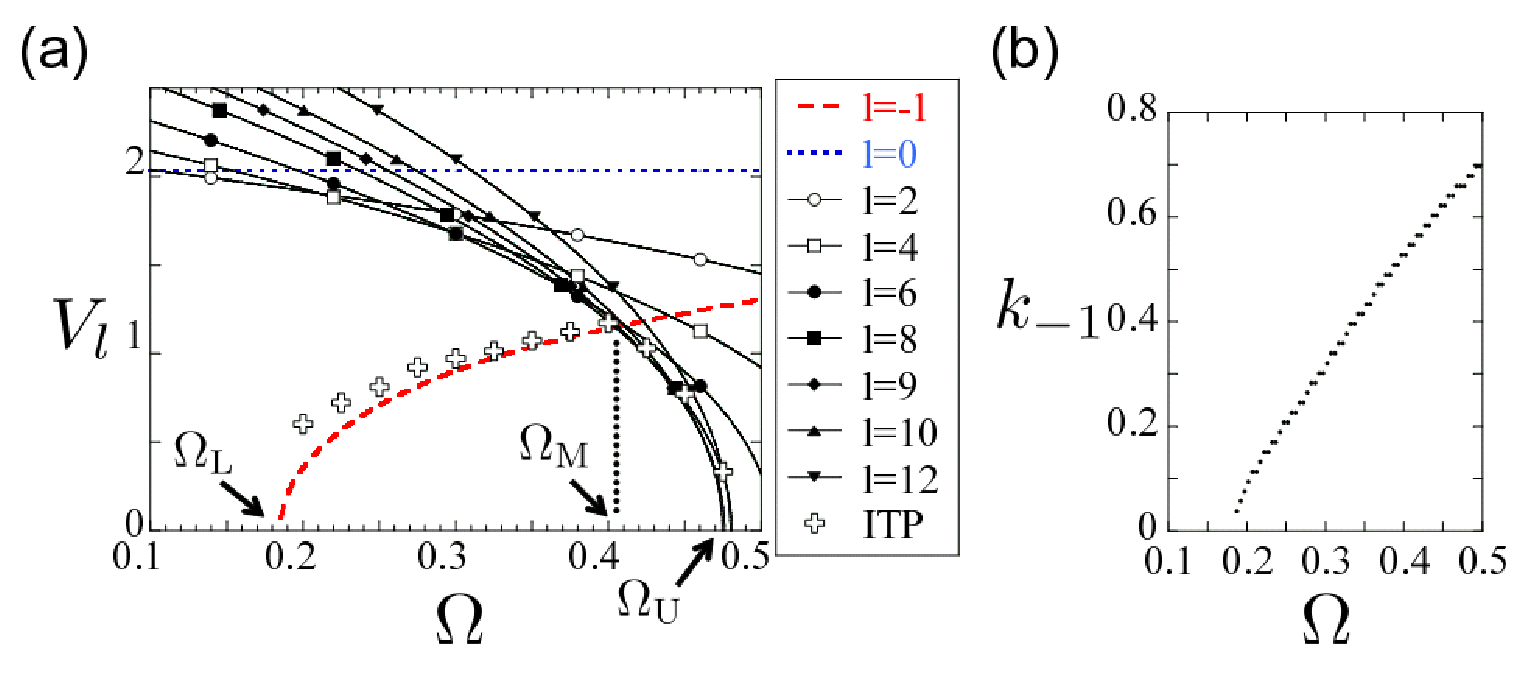}
  \caption{
 (a) Critical velocities $V_l(\Omega)$ obtained by the BdG analysis for the single-vortex state with $g_{\rm 2D} =  500$.
 The dashed, dotted, and solid lines show $V_l(\Omega)$ for Kelvin waves ($l=-1$), varicose waves ($l=0$), and several surface waves ($l>0$), respectively.
 The cross marks show the minimum critical velocity $V_c=\min_l(V_l)$ obtained by the ITP analysis.
 (b) Critical wave number $k_{-1}$ of the Kelvin waves excited at $V=V_{-1}$.
}
\label{fig:V_l}
\end{figure}

 To observe the DG instability, the condition $V_{-1} < V_l~(l \neq -1)$ must be satisfied, because the single-vortex states become unstable due to the Landau instability exciting the other modes if $V_{l \neq -1} <V< V_{-1}$.
 Figure \ref{fig:V_l} shows the critical velocity $V_l$ for various $l$ and the critical wave number $k_{-1}$ obtained by Eq. (\ref{eq:V_l}) with the result of Fig. \ref{fig:spect}(c).
 The critical velocities for surface waves with $l<-1$ are always higher than $V_{-1}$, not shown in Fig. \ref{fig:V_l}(a).
 The critical velocity $V_0$ for varicose waves is typically higher than $V_{-1}$.
 This is because the dispersion relation of varicose waves is phonon-like while that of Kelvin waves is quadratic for small wave numbers [see Fig. \ref{fig:spect}(c)].
 While the critical velocity $V_{-1}$ for Kelvin waves monotonically increases with $\Omega$, $V_{l>0}$ for surface waves decreases.
 The increase in $\Omega$ reduces $V_{l>0}$ for some $l$ to be equal to $V_{-1}$ at $\Omega=\Omega_{M}$.
 Thus the DG instability can occur in the region $\Omega_{L} < \Omega < \Omega_{M}$.
 This region generally appears because the condition $\Omega_{L} < \Omega_{U}$ is satisfied \cite{IsoshimaCritical} and $\Omega_{M}$ is always placed between $\Omega_{L}$ and $\Omega_{U}$ for positive $g$.

\subsection{Kelvin wave amplification}
 Next, we discuss the amplification of Kelvin waves after spontaneous radiation due to the Landau instability.
 In order to understand the amplification process qualitatively, we include the dissipation term phenomenologically \cite{Kasamatsu} in the left side of Eq. (\ref{eq:GP}),
\begin{eqnarray}
{\textstyle
(i-\gamma)\frac{\partial}{\partial t} \Psi= ( -\frac{\nabla^2}{2}+\frac{\rho^2}{2}+g|\Psi|^2 -\mu-V\hat{p}_z -\Omega\hat{l}_z) \Psi,
}
\label{eq:GP_gamma}
\end{eqnarray}
 where $\gamma (>0)$ is the dissipation term.
 The value of $\gamma$ should be determined from the interaction between an external environment and the condensate.
 The role of the environment, which dissipates the thermodynamic energy of Eq. (\ref{eq:Kpotential}) and thus causes the Landau instability, can be played by an external potential or thermal cloud agitated by the potential \cite{Raman_stir,Kasamatsu,VcBEC-BCS}.
 The helical motion of the external potential can be realized in experiment. For example, a helically moving optical lattice can be made by rotating the sources of two counterpropagating laser beams with different frequencies.

 Because of the dissipation term in Eq. (\ref{eq:GP_gamma}),
 Kelvin waves and surface waves are amplified when $V$ exceeds their critical velocity for $\Omega_{L}<\Omega<\Omega_{M}$ and $\Omega_{M}<\Omega<\Omega_{U}$, respectively.
 Figures \ref{fig:spect}(a) and \ref{fig:spect}(b) show the typical constant surface densities $|\Psi|^2$ of condensates at initial stages in the amplification dynamics obtained by numerically solving Eq. (\ref{eq:GP_gamma}).
 The surface wave amplification causes a growing ripple which propagates helically on the surface of the condensate [Fig. \ref{fig:spect}(b)], which leads to dragging more vortices into the condensate.
 The Kelvin wave amplification makes a helical configuration of the vortex line [Fig. \ref{fig:spect}(a)].
 Accompanying the Kelvin wave amplification, the radius of the helix is increased with time, which is qualitatively the same as the dynamics of the DG instability in superfluid helium.
 This amplification enables us to observe the Kelvin waves \cite{KelvinExp} and their wave numbers $k \sim k_{-1}$ [Fig. \ref{fig:V_l}(b)] when $V \sim V_{-1}$.
 From this observation, we can get the dispersion relation $\omega^0_{k_{-1},-1}=k_{-1}V_{-1}-\Omega$ of Kelvin waves through the relation Eq. (\ref{eq:V_l}), which has never been experimentally obtained to date.

\subsection{Critical velocity for the multivortex states}
 It is also interesting to study the DG instability in the presence of more than one vortex.
 In multivortex states, the rotational symmetry of the condensate is broken and the excited states are no longer eigenstates of the angular momentum,
 which makes the above analysis of the BdG equations very complicated.
 In this case, for the purpose of investigating the critical velocity of the DG instability,
 it is convenient to follow a numerical procedure using the imaginary time propagation (ITP) of the GP Eq. (\ref{eq:GP}).
 Although the ITP is employed to find the state with a minimum of $K(V,\Omega)$,
 we use it with the following procedure.
 First, a vortex state with $n_v$ ($>1$) vortices is obtained with the ITP for $V=0$ and a fixed value $\Omega$,
 where each multivortex state is stable within a certain range of $\Omega$.
 Second, after increasing $V$ a little and adding a small random noise to the multivortex state,
 the ITP is restarted.
 If the initial multivortex state is kept,
 the second step is repeated.
 If not, the velocity $V$ at that time is regarded as the critical velocity $V_c$.
 We plot the ITP results of $V_c$ for the single-vortex states as cross marks in Fig. \ref{fig:V_l}(a), compared to the results of the BdG analysis.
 Both results are consistent, which means that the ITP analysis could be applicable to this problem.

\begin{figure} [htbp] \centering
  \includegraphics[width=1. \linewidth]{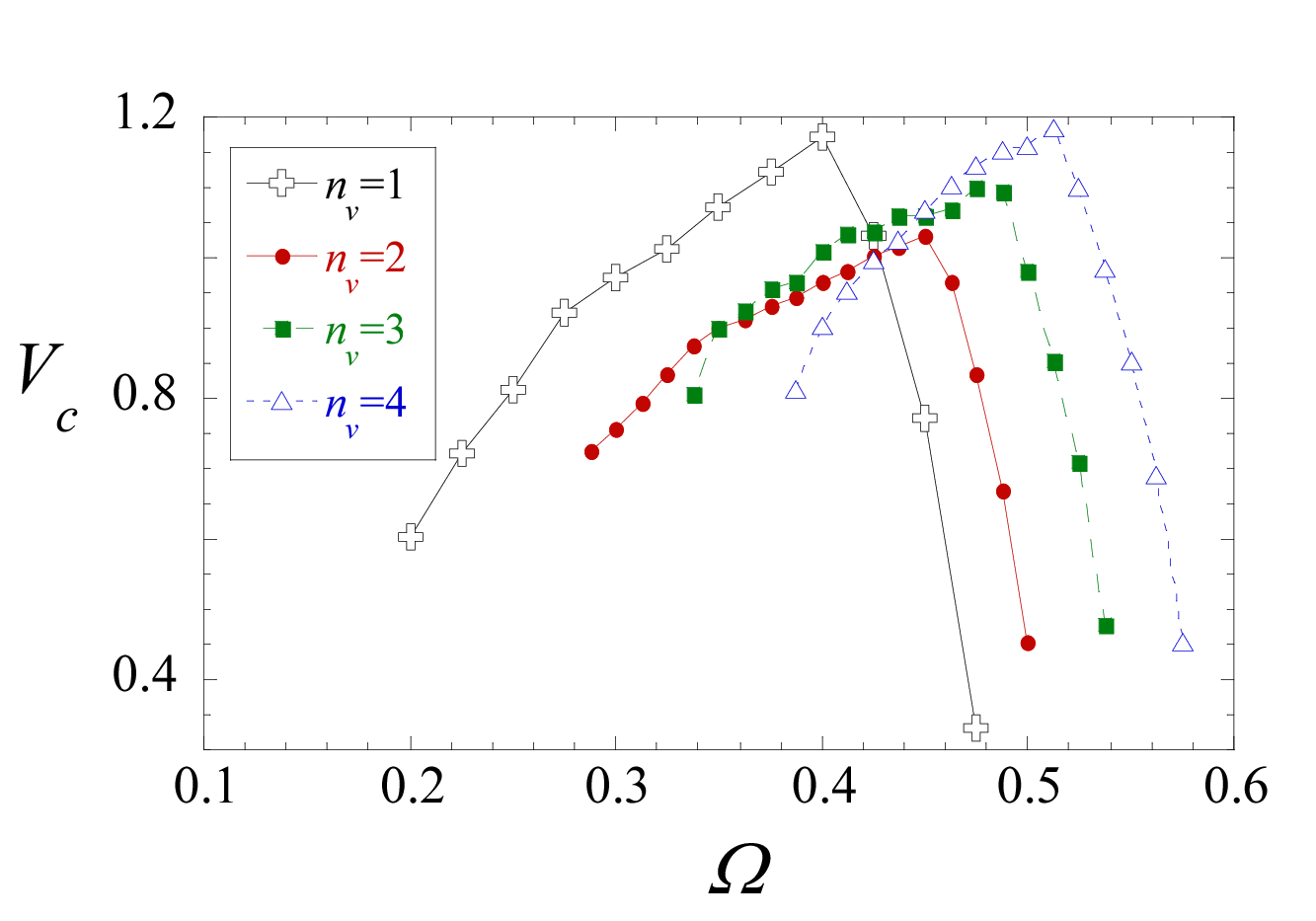}
  \caption{
 The smallest critical velocity $V_c(\Omega)$ for vortex states with $n_v(=1,2,3,4)$ vortices and $g_{\rm 2D} = 500$ obtained by the ITP analysis.
 The critical velocities for the Kelvin and the surface waves cross at each maximum.
 Thus the Kelvin and the surface waves are amplified to the left and right sides of each maximum, respectively.
}
\label{fig:v234}
\end{figure}
 We numerically investigate how $V_c$ depends on $\Omega$ and which modes are excited via the instability for multivortex states with $g_{\rm 2D}=500$ (Fig. \ref{fig:v234}).
 The critical velocities for the vortex number $n_v = 2, 3, 4$ behave similar to that of $n_v = 1$;
 the critical velocities for the Kelvin and the surface waves cross at each maximum in Fig. \ref{fig:v234}.
 In fact, the Kelvin and the surface waves are amplified to the left and right sides of each maximum, respectively.
 However, there are quantitative differences between them.
 The critical velocity for surface waves is increased with the number $n_v$ in Fig. \ref{fig:v234}.
 This is because, for a fixed $\Omega$ it is easier to increase the angular momentum of the condensate for a vortex state with small $n_v$ than for one with large $n'_v$ ($n'_v > n_v$).
 On the other hand, there is a quantitative difference between the critical velocity for the Kelvin wave of the single-vortex and those of the three multivortex states ($n_v=2, 3, 4$).
 This fact can be explained in the following.
 Since the Kelvin modes disturb locally the density around the vortex core for the single-vortex state,
 the bulk energy $g|\Psi({\bm r})|^2$, which is the local interaction energy of the bulk around the core, should be dominant to the dispersion relation $\omega^0_{k,-1}$ of Kelvin waves.
 It is useful to consider the facts that $\omega_{0,-1}=-\Omega_{L}$ is increased with $g_{\rm 2D}$ \cite{IsoshimaCritical} and the bulk energy is typically increased with $g_{\rm 2D}$.
 As a result, the critical velocity typically increases with the bulk energy.
 This consideration would be applicable to the multivortex states with a few vortices,
 where the intervortex separation is large enough that the interaction between vortices is negligible.
 But there is qualitative difference between the single-vortex state ($n_v=1$) and the three multivortex states ($n_v=2,3,4$);
 a vortex is centered for the former although vortices are off centered for the latter.
 Since the bulk energy $g|\Psi({\bm r})|^2$ is decreased with $\rho$,
 the critical velocity of the Kelvin waves for the single-vortex state is larger than those for the multivortex states.
 However, the critical velocities for the three multivortex states ($n_v=2, 3, 4$) are relatively close to each other.
 This is because the three multivortex states are not so different in the bulk energy.

\subsection{Vortex reconnections}
\begin{figure} [bthp] \centering
  \includegraphics[width=1. \linewidth]{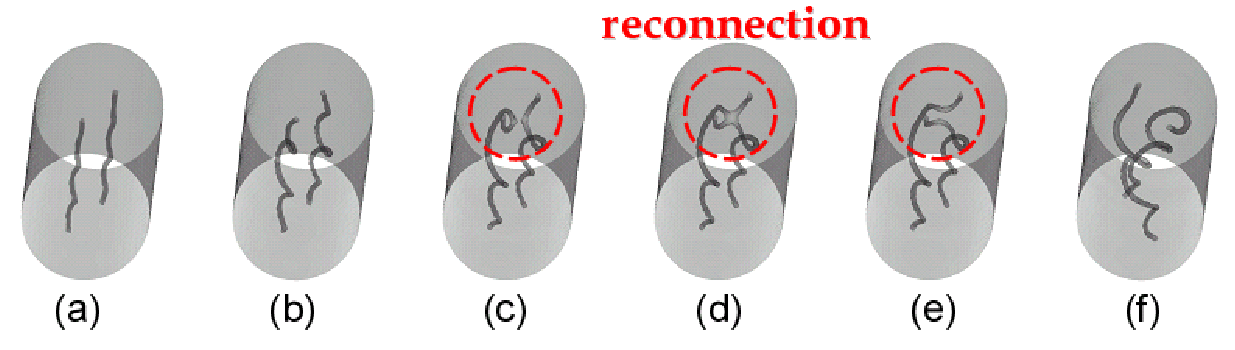}
  \caption{
 Nonlinear vortex dynamics caused by the DG instability in a trapped BEC with two vortices. [(a) and (b)]
 After Kelvin waves are amplified, [(c)-(e)] vortex reconnection takes place, (f) leading to complex dynamics.
 The values of the parameters are $g_{\rm 2D}=500$, $L=62.2768$, $\Omega=0.35$, $V=1.00$, and $\gamma=0.05$.
 The times are (a) $t=0$, (b) $147.96$, (c) $179.52$, (d) $181.50$, (e) $182.48$, and (f) $241.67$.
}
\label{fig:dyv2}
\end{figure}
 When the Kelvin waves are amplified to the order of the distance between adjoining vortices, nonlinear effects (vortex interactions and vortex reconnections) start to work.
 Figure \ref{fig:dyv2} shows the dynamics of a two-vortex state obtained by numerically solving Eq. (\ref{eq:GP_gamma}).
 The amplification of Kelvin waves [Figs. \ref{fig:dyv2}(a) and \ref{fig:dyv2}(b)] causes a vortex reconnection [Figs. \ref{fig:dyv2}(c)-\ref{fig:dyv2}(e)].
 Then amplification of the Kelvin waves continues [Fig. \ref{fig:dyv2}(f)], leading to complex dynamics.
 In this way, the vortex reconnections are controllable with the DG instability in atomic BECs.
 This enables us to investigate the vortex reconnection together with the acoustic emission \cite{Ogawa} in atomic BECs,
 which is impossible in superfluid helium because of its incompressibility.

\subsection{Transition from a vortex lattice to a vortex tangle}
\begin{figure} [htbp] \centering
  \includegraphics[width=1. \linewidth]{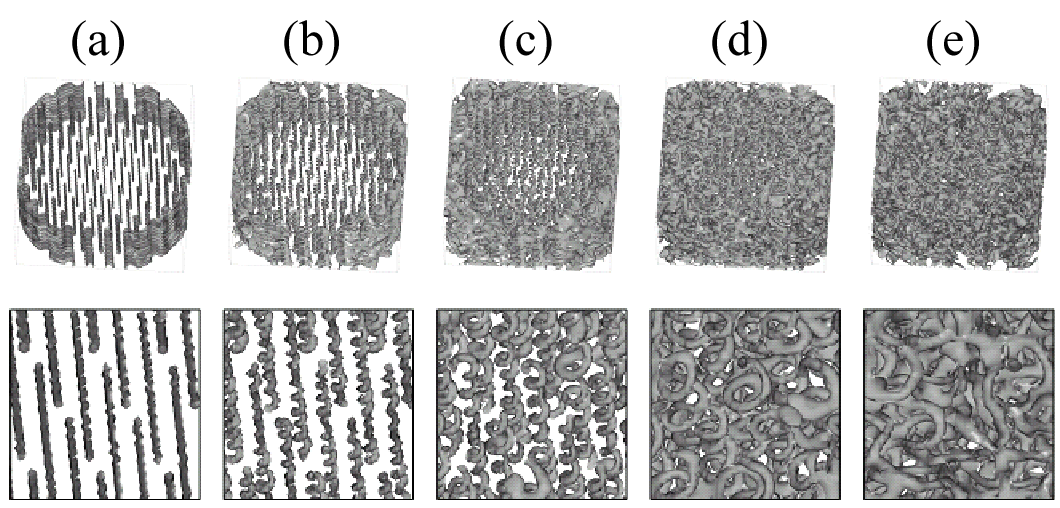}
  \caption{
Transition dynamics  (a) from a vortex lattice to (e) a vortex tangle due to the DG instability in a trapped BEC.
 Figures in the lower row show the magnified images in the center region of upper ones.
 [(a)-(c)] Kelvin wave is amplified on each vortex.
 (d) Then a lot of vortex reconnections start to take place with adjoining vortices,
 (e) which results in a dense vortex tangle.
 The values of the parameters are $g_{\rm 2D}=9000$, $L=10.02$, $\Omega=0.8$, $V=6.95$, and $\gamma=0.05$.
 The times are (a) $t=0$, (b) $1.02$, (c) $1.53$, (d) $2.04$, and (e) $2.55$.
}
\label{fig:lattice}
\end{figure}
 It is expected that if there are many vortices in the initial states, the vortex reconnection occur frequently with the neighboring vortices.
 Then the vortex dynamics will get much complicated due to the strong nonlinear effects.
 As discussed for superfluid helium \cite{Tsubota_Araki}, the DG instability in vortex lattices can lead to a dense vortex tangle, namely quantum turbulence.
 A similar scenario is expected for atomic BECs.
 In fact, we numerically obtained a transition from a vortex lattice [Fig. \ref{fig:lattice}(a)] to a dense vortex tangle [Fig. \ref{fig:lattice}(e)], due to a lot of vortex reconnections [Fig. \ref{fig:lattice}(d)] after the amplification of Kelvin waves on each vortex [Fig. \ref{fig:lattice}(a)-\ref{fig:lattice}(c)].
 Thus, the DG instability gives the possibility of realizing quantum turbulence in atomic BECs \cite{Kobayashi}.

\section{CONCLUSION}
 In conclusion, we theoretically discussed a different type of Landau instability in atomic BECs driven by a helically moving environment.
 In particular, we studied the Landau instability of Kelvin waves, which generalized the DG instability from the thermodynamic point of view.
 The DG instability is possible for both single-vortex and multivortex states.
 These phenomena lead to the direct observation of the dispersion relation of Kelvin waves, vortex reconnections, and quantum turbulence.

\begin{acknowledgments}
 H.T. acknowledges the support of a Grant-in-Aid for JSPS Fellows (Grant No. 199748).
 K.K. acknowledges the supports of Grant-in-Aid for Scientific Research from JSPS (Grant No. 18740213).
 M.T. acknowledges the support of a Grant-in-Aid for Scientific Research from JSPS (Grant No. 18340109) and a Grant-in-Aid for Scientific Research on Priority Areas from MEXT (Grant No. 17071008).
\end{acknowledgments}

\end{document}